%% file: IFA009-article.tex
\documentclass[epj]{svjour}

\usepackage[utf8]{inputenc} %
\usepackage[T1]{fontenc}
\usepackage{graphicx}
\usepackage{isotope}
\usepackage{siunitx}
\usepackage[draft]{fixme}
\usepackage{bm}
\usepackage{xcolor}
\usepackage{xspace}
\usepackage{ulem}
\usepackage{booktabs}

\graphicspath{{figures/}}
\usepackage{color}

\newcommand{\Boron}{\isotope[11]{B}}
\newcommand{\Be}{\isotope[8]{Be}}
\newcommand{\Carbon}{\isotope[12]{C}}
\newcommand{\unsim}{\mathord{\sim}}
\newcommand{\etal}{\textit{et al.}\xspace}

\usepackage[switch]{lineno}
\modulolinenumbers[1]
\usepackage{hyperref}
\usepackage{cleveref}

\begin{document}
\title{Resolving the \textsuperscript{11}B$\bm{(p,\alpha_{0})}$
  Cross Section Discrepancies between 0.5 and 3.5 MeV}

\author{
  Michael Munch\inst{1}
  and Oliver Sølund Kirsebom\inst{1}
  and Jacobus Andreas Swartz\inst{1}%
  \thanks{Present address: Center for Nuclear Technologies, Technical University of Denmark,
    Frederiksborgvej 399, 4000 Roskilde, Denmark}
  and Hans Otto Uldall Fynbo\inst{1}
}

\authorrunning{M. Munch \textit{et al.}}
\titlerunning{Resolving the \textsuperscript{11}B$(p,\alpha_{0})$
  Cross Section Discrepancies between 0.5 and \SI{3.5}{\MeV}}

\offprints{H. O. U. Fynbo (\texttt{fynbo@phys.au.dk})}          %
\institute{Department of Physics and Astronomy, Aarhus University, Ny Munkegade 120, 8000
  Aarhus C, Denmark}
\date{Received: date / Revised version: date}
\input{IFA009-abstract}
\maketitle

\input{IFA009-introduction}
\input{IFA009-experiment}

\input{IFA009-analysis}
\input{IFA009-discussion}

\input{IFA009-conclusion}

\input{IFA009-acknowledge}

\bibliography{IFA009-paper}

\end{document}

%% file: IFA009-abstract.tex
\abstract{
The reaction $\Boron(p,3\alpha)$ is relevant for fields as diverse as material science, nuclear
structure, nuclear astrophysics, and fusion science. However, for the channel proceeding via
the ground state of \Be, the available cross-section data shows large discrepancies of both normalization and
energy scale.
The present paper reports on a measurement of the $\Boron(p,\alpha_{0})$ cross section using an
array of modern large area segmented silicon detectors and low beam current on an enriched thin target with the aim of
resolving the discrepancies amongst previous measurements.
  \PACS{
    {PACS-key}{discribing text of that key}   \and
    {PACS-key}{discribing text of that key}
  } %
} %

%% file: IFA009-introduction.tex
\section{Introduction}
\label{sec:intro}

The reaction $\Boron(p,3\alpha)$ 
is relevant for fields as diverse as material science, nuclear structure, nuclear astrophysics,
and fusion science. In a sequential picture, this reaction can proceed via the ground state of
\Be, $\Boron(p,\alpha_0)\Be$, or via the first excited state, $\Boron(p,\alpha_1)\Be^{*}$.

Boron depth profiling in bulk matter can be studied using the technique of nuclear reaction
analysis (NRA) where a nuclear reaction on the element of interest produces a particle with
high enough energy to escape the material, and the energy distribution of the emitted particles
can be used to deduce the depth profile.  Boron depth profiling using NRA requires an accurate
cross section of the $\Boron(p,\alpha_0)$
reaction in the energy range up to \SIrange{3}{4}{\MeV}~\cite{Kokkoris2010}.

The rate of the $\Boron(p,3\alpha)$
reaction is also relevant for understanding the astrophysical abundances of the light elements
Li, Be, and B~\cite{Boesgaard2005,Lamia2012}. Specifically, the abundances of these elements in
stellar atmospheres can be depleted due to plasma mixing phenomena inside the stars at a rate
which depends on fundamental nuclear cross sections.

The $\Boron(p,3\alpha)$
reaction is considered a candidate for fusion energy generators as an alternative to e.g. the
most favored $t(d,n)\alpha$
reaction~\cite{Moreau1977}. In order to evaluate the feasibility of the $\Boron(p,3\alpha)$
reaction accurate cross sections are required~\cite{Sikora2016}.

Finally, the cross section of the $\Boron(p,3\alpha)$ reaction
provides a direct probe for resonances in \Carbon{} above the \Boron$+p$
threshold. 
For proton energies up to \SI{2}{\MeV} the excitation spectrum exhibits resonances due to three
isospin $T=1$ states at 16.11, 16.62 and \SI{17.76}{\MeV}, which have spin and parity of $2^+$, $2^-$ and $0^+$ and whose properties are well established. Additionally, a very broad $1^-$, $T=1$ resonance and sub-threshold resonances of unknown spin and parity are believed to contribute to the excitation spectrum in the region between the 2$^-$ and 0$^+$ resonances~\cite{Symons1963,Segel1965,Barker2002}.

\begin{figure}[b]
  \centering
  \includegraphics[width=\columnwidth]{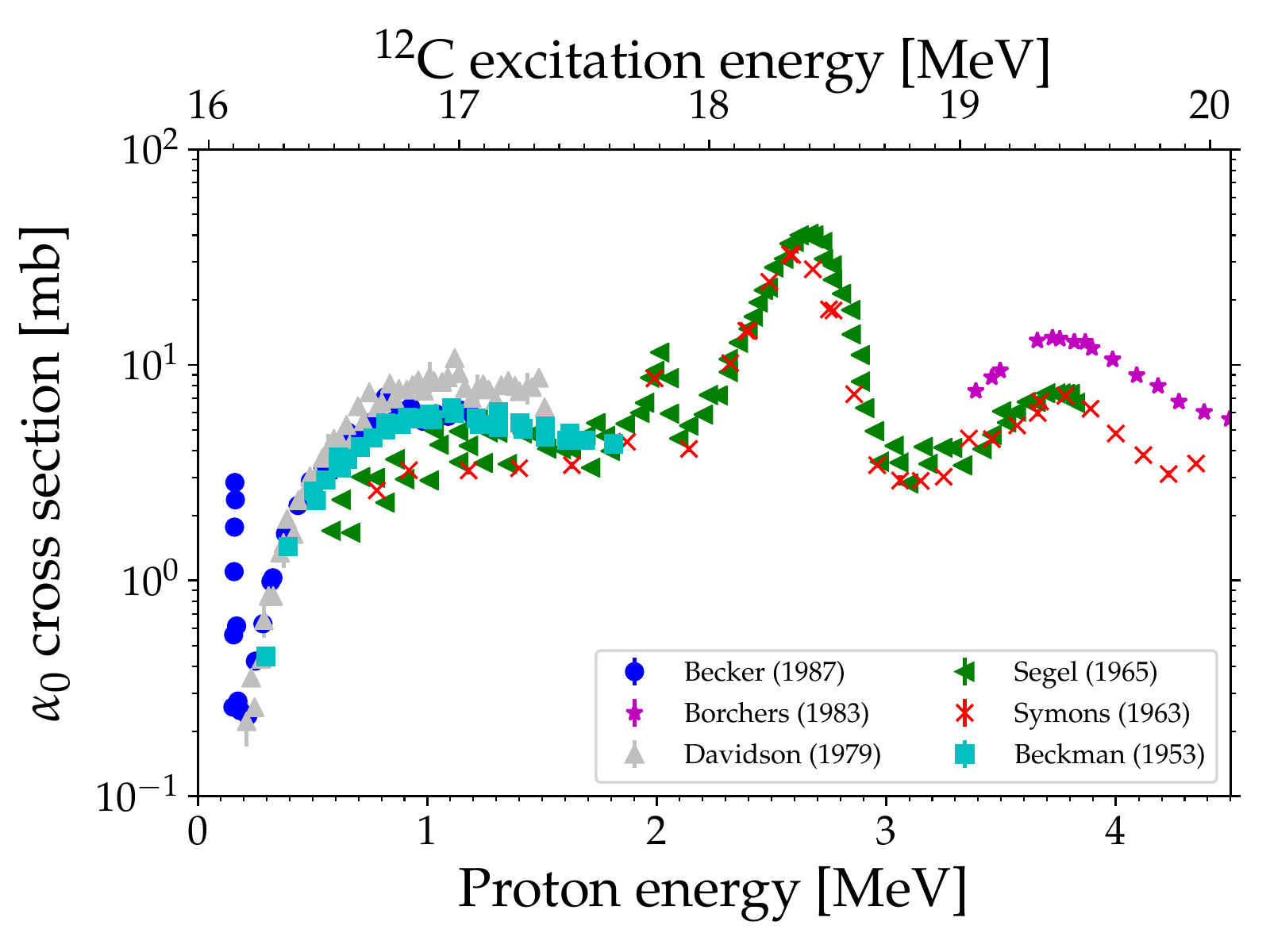}
  \caption{%
    Angle-integrated $\alpha_{0}$
    cross section as reported by
    Refs.~\cite{Beckman1953,Symons1963,Segel1965,Davidson1979,Borchers1983,Becker1987}
    (tabulated data available via EXFOR~\cite{OTUKA2014272}).
    Deviations both in the absolute normalization and energy calibration are evident. 
  }
  \label{fig:a0-unscaled}
\end{figure}

A handful of measurements over extended ranges of beam energies and/or emission angles of the
final state particles have been published, but these measurements differ significantly both in
the energy scale and in the absolute normalization of the cross
sections. \Cref{fig:a0-unscaled} shows cross section data from a set of measurements of the
$\Boron(p,\alpha_0)$ reaction which is the focus of the present paper. As is evident, there are significant
deviations of the order of \SI{100}{\keV} in the energy scale and up to \SI{70}{\%} in the magnitude of the
cross section between different measurements. Likely sources of such deviations can be accelerator
energy calibrations and inaccurate evaluations of target thicknesses.

Some of us have recently published a re-measurement of the $\Boron(p,3\alpha)$ reaction at a proton 
energy corresponding to the \SI{16.11}{MeV} $2^+$ $T=1$ resonance, which produced a more precise 
and accurate determination of the cross section at that energy~\cite{Munch2018}. Here we use the same 
experimental approach to study the $\Boron(p,\alpha_0)$ reaction for proton energies in the range 
\SIrange{0.5}{3.5}{\MeV}.

%% file: IFA009-experiment.tex
\begin{figure}[b]
  \centering
  \includegraphics[width=\columnwidth]{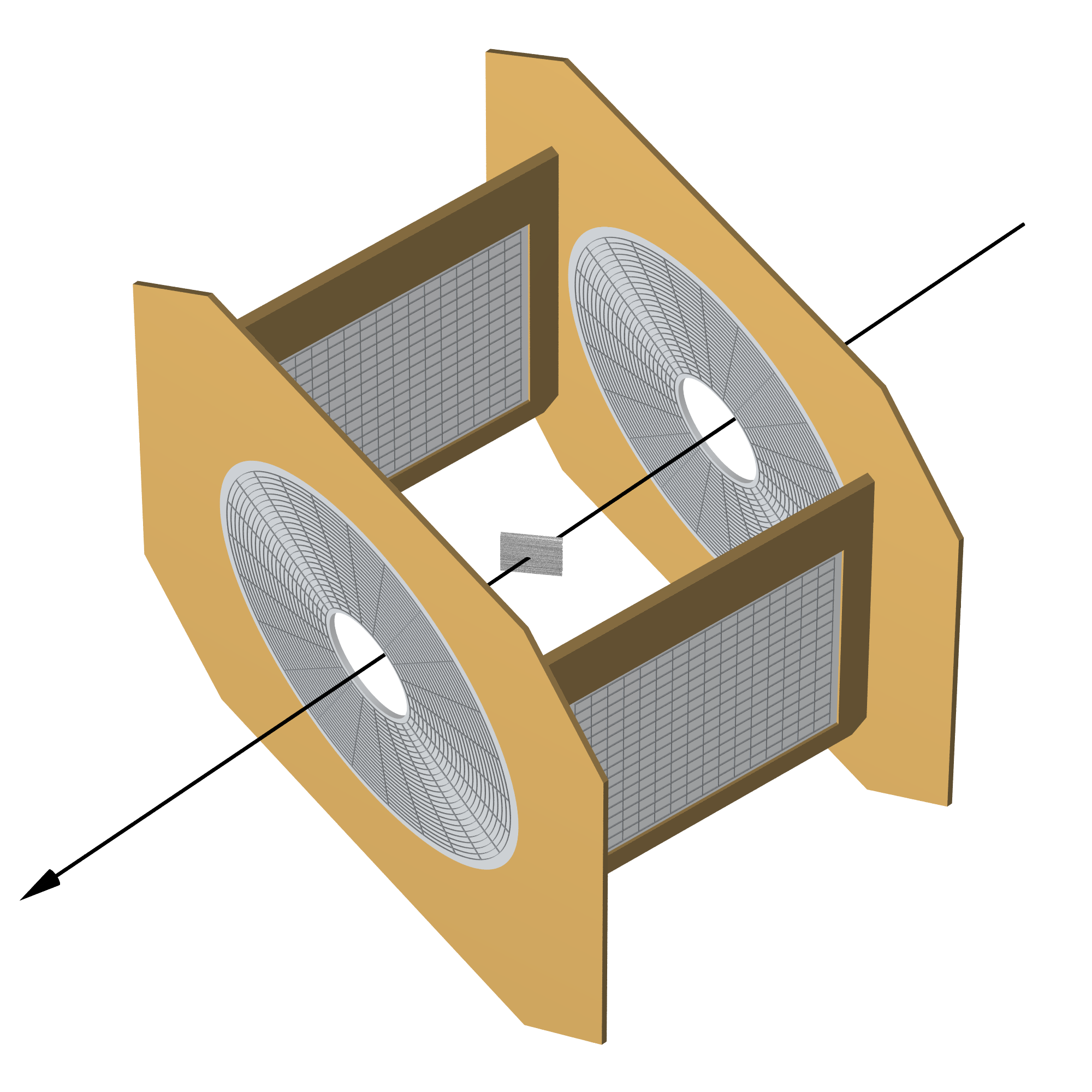}
  \caption{Detection setup. Two annular detectors were placed \SI{42}{\mm} up- and downstream of
    the target, while two quadratic detectors were placed at \SI{85}{\degree} and
    \SI{95}{\degree}. The target was oriented \SI{45}{\degree} with respect to the beam axis,
    which is indicated by the arrow.}
  \label{fig:setup}
\end{figure}

\section{Experiment}
\label{sec:experiment}

The experiment was conducted at the Aarhus University \SI{5}{\MeV} accelerator
which provided a beam of protons with an energy between \num{0.5} and \SI{3.5}{\MeV}.  
The beam impinged on a thin boron target which was oriented \SI{45}{\degree} with respect to
the beam axis. The beam current was measured \SI{1}{\m} downstream of the target position and
the beam spot was collimated to be $1\si{\mm}\times1$\si{\mm} by a set of horizontal and vertical slits.

The boron target was manufactured by evaporating \SI{99}{\%} enriched \Boron{} onto a thin
$\unsim \SI{4}{\ug\per\cm\squared}$ natural carbon backing. In order to determine the target
thickness the target was oriented at angles of \SI{0}{\degree} and \SI{180}{\degree} with respect to the beam axis
and bombarded with \SI{2}{\MeV} $\alpha$ particles. The \Boron{} thickness could then be deduced from
the energy shift of the carbon peak using the procedure described in Ref.~\cite{Munch2018}. The
result was \SI{12.6(12)}{\ug\per\cm\squared}. 

Charged particles were detected using an array of four Double Sided Silicon Strip Detectors (DSSD) in
close geometry%
.  These provided a simultaneous detection of energy and position of emitted charged
particles. The array consisted of two annular DSSDs (S3 from Micron Semiconductors) placed
\SI{42}{\mm} up- and downstream of the target.  Additionally, two quadratic DSSDs (W1 from
Micron Semiconductors) were placed approximately \SI{36}{\mm} from the target center at an
angles of \SI{85}{\degree} and \SI{95}{\degree} with respect to the beam axis.

The trigger logic was handled by a \textsc{GSI Vulom4b} module \cite{VULOM4b} running the
\textsc{TRLO II} firmware \cite{Johansson2013}. The trigger logic was configured as a logical
OR between all four detectors, with the downstream detector downscaled by a factor of eight. The
firmware maintains scaler values for the accepted and total number of triggers for each
detector. The deadtime for each run was determined from these.

The energy scan was performed in steps of \SI{100}{\keV} with smaller steps near well-known
resonances. Data were acquired at each energy for approximately one hour with a beam
current of 0.3--1~nA. Longer measurements were performed for the $0^{+}$ state at \SI{17.76}{\MeV} and the $3^{-}$ state at \SI{18.35}{\MeV}. 
The result of these longer measurements will be reported separately \cite{Kirsebom2018}.

%% file: IFA009-analysis.tex
\begin{figure}[b!]
  \centering
  \includegraphics[width=\columnwidth]{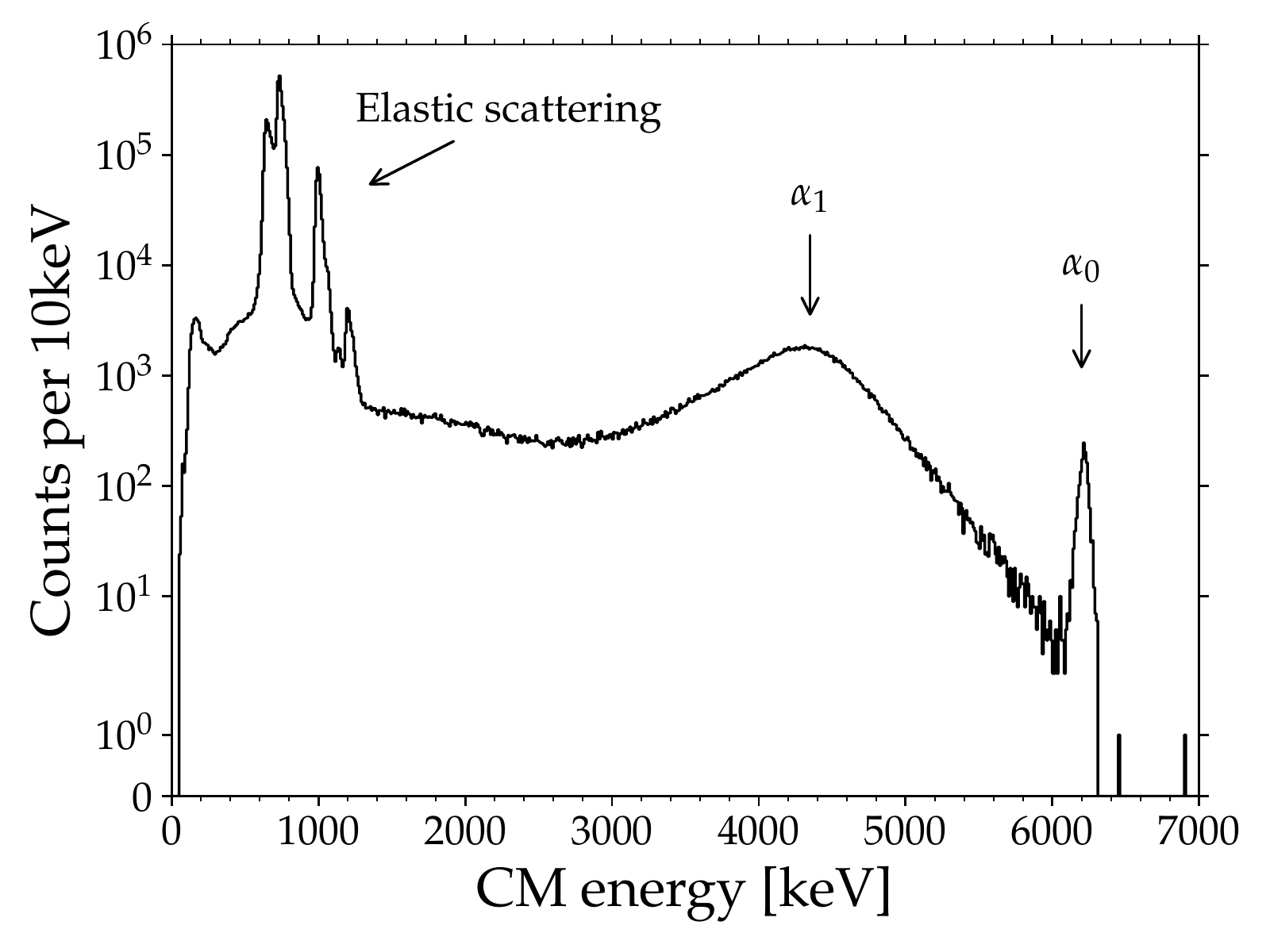}
  \caption{Center-of-mass energy distribution for a proton energy of \SI{800}{\keV}. At
    high energy, one sees a sharp peak corresponding to $\alpha_{0}$.
    Below this peak is a broad distribution corresponding to reactions proceeding via the
    $\alpha_{1}$
    channel.  Around \SI{1}{\MeV} there are various peaks corresponding to elastic proton
    scattering. The energy of the elastic protons might be larger than the beam energy
    since they have been corrected for $\alpha$ particle energy loss.  
  }
  \label{fig:cm-energy}
\end{figure}

\begin{figure*}[p]
  \begin{minipage}{1.0\linewidth}
    \centering
    \includegraphics[width=\columnwidth]{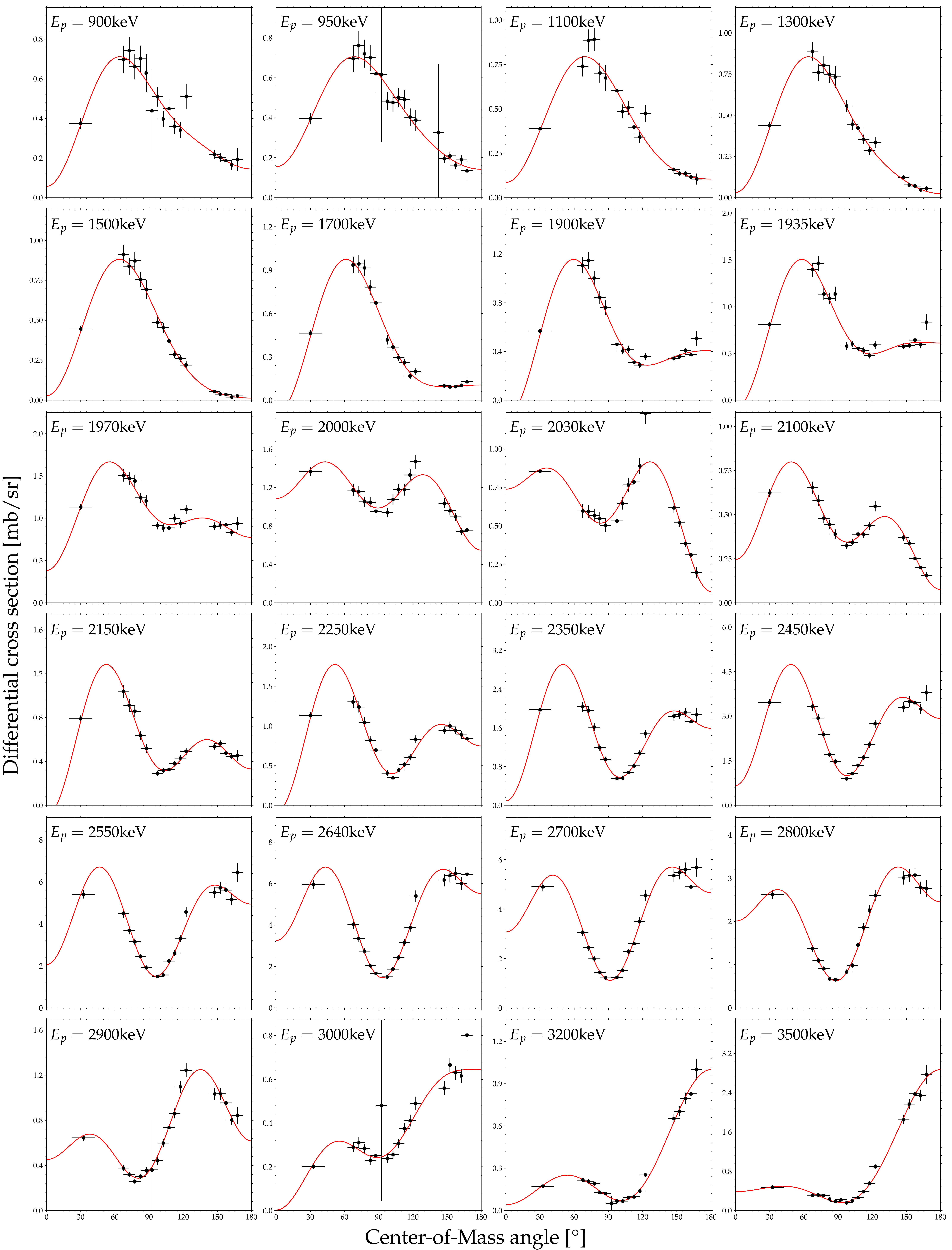}
    \caption{
      Selection of angular distribution from 900 to \SI{3500}{\keV}.
      A slow evulotion from slightly forward focussed below \SI{2}{\MeV}
      to approximately symmetric around \SI{2.4}{\MeV}
      to strongly backward focussed above \SI{3}{\MeV}.
    }
    \label{fig:angular-dist}
  \end{minipage}
\end{figure*}

\section{Data reduction}
\label{sec:analysis}

\begin{figure}[!b]
  \centering
  \includegraphics[width=\columnwidth]{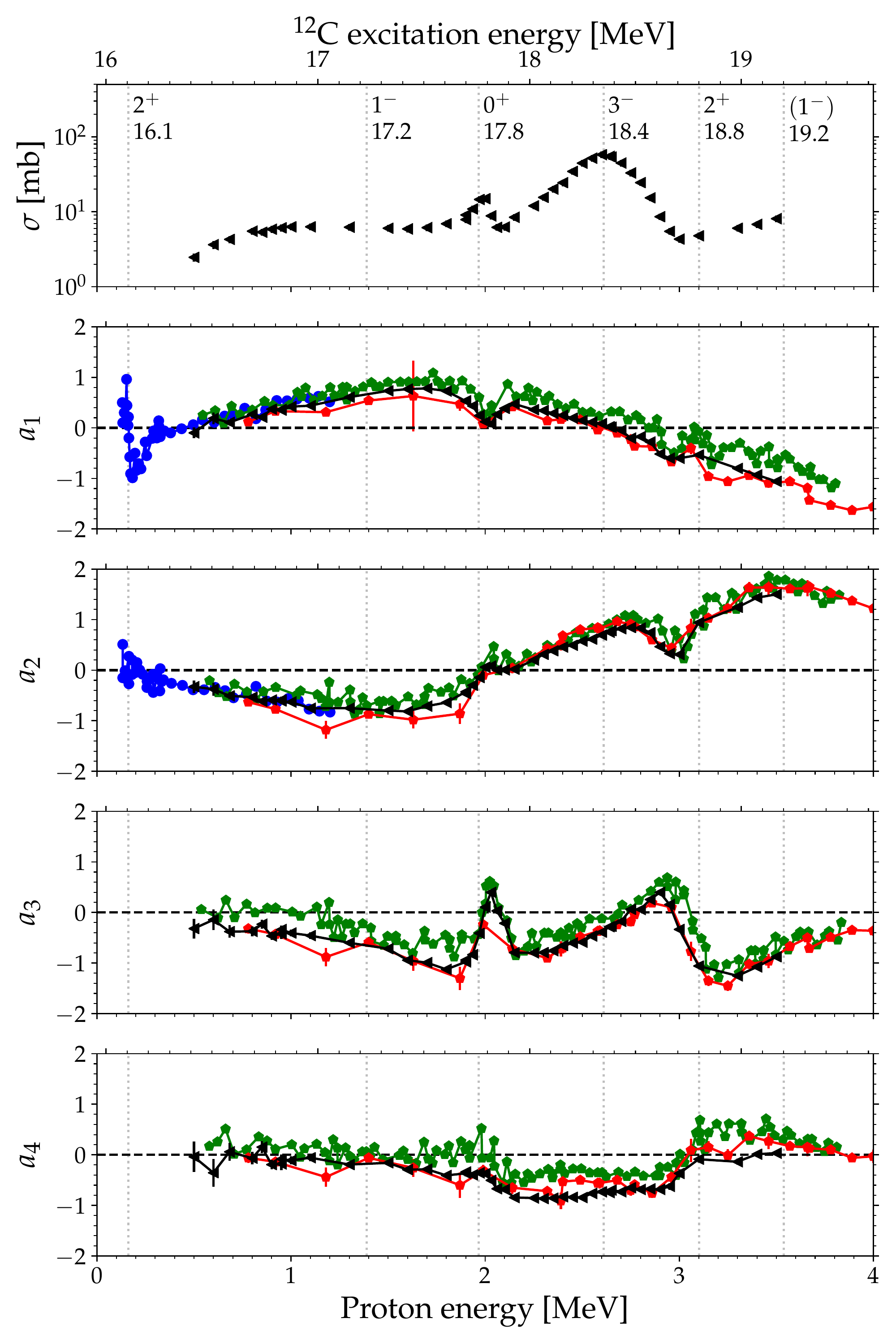}
  \includegraphics[width=\columnwidth]{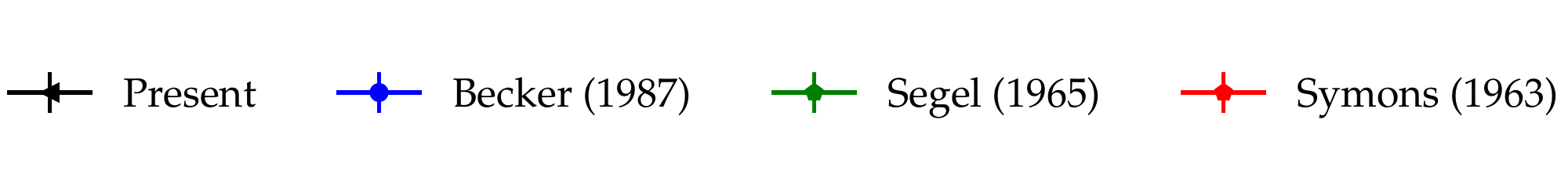}
  \caption{Angle-integrated cross section and Legendre coefficients shown as a function of beam
    energy together with the coefficients obtained by
    Refs.~\cite{Symons1963,Segel1965,Becker1987}. The dashed vertical lines indicate the
    position of natural parity states in \Carbon{} as listed in Ref.~\cite{Kelley2017}.
  }
  \label{fig:as}
\end{figure}

\begin{table}[b]
  \centering
  \caption{ Tabulated angle-integrated cross sections. The uncertainty is approximately
    \SI{10}{\%} for all data points except the lowest two where it is \SI{12}{\%}.  }
  \label{tab:cross}
  \begin{tabular*}{\columnwidth}{
    @{\extracolsep{\fill}}
    S[table-format =  4.0]
    S[table-format =  2.1(2)]
    S[table-format =  4.0]
    S[table-format =  2.1(2)]
    }
    \toprule 
    {$E_{p}$ [keV]} & {$\sigma_{\alpha0}$ [mb]} & {$E_{p}$ [keV]} & {$\sigma_{\alpha0}$ [mb]}   \\
    \midrule
    500 & 2.5(3) & 2150 & 8.4(9) \\
    600 & 3.6(4) & 2250 & 12.0(12) \\
    683 & 4.3(5) & 2300 & 15.5(16) \\
    800 & 5.5(6) & 2350 & 20(2) \\
    850 & 5.3(6) & 2400 & 24(2) \\
    900 & 5.8(6) & 2450 & 35(3) \\
    950 & 6.1(6) & 2500 & 44(4) \\
    950 & 6.1(6) & 2550 & 52(5) \\
    1000 & 6.3(6) & 2600 & 58(6) \\
    1100 & 6.3(6) & 2640 & 55(6) \\
    1300 & 6.2(6) & 2650 & 54(5) \\
    1500 & 6.0(6) & 2700 & 45(5) \\
    1600 & 5.9(6) & 2750 & 33(3) \\
    1700 & 6.1(6) & 2800 & 24(2) \\
    1800 & 6.9(7) & 2850 & 15.3(15) \\
    1900 & 7.9(8) & 2900 & 8.6(9) \\
    1900 & 9.1(9) & 2950 & 5.5(6) \\
    1935 & 10.9(11) & 3000 & 4.3(4) \\
    1970 & 14.6(15) & 3100 & 4.8(5) \\
    2000 & 14.9(15) & 3300 & 6.0(6) \\
    2030 & 8.8(9) & 3400 & 6.8(7) \\
    2060 & 6.2(6) & 3500 & 8.1(8) \\
    2100 & 6.3(6) \\
    \bottomrule   
  \end{tabular*}
\end{table}

The analysis was performed event by event on list-mode type data. The
detected particles were assumed to be $\alpha$ particles and the
energies were corrected for losses in the target foil and the dead layers of the detectors, which are $\unsim \SI{500}{\nm}$ thick for the annular design and
$\unsim \SI{100}{\nm}$ for the quadratic design.  The corrected energies and
the angles were transformed to the beam-target center-of-mass (CM) system. An
example of the resulting CM energy spectrum can be seen in
\cref{fig:cm-energy} with a clear peak at high energy corresponding to
the $\alpha_{0}$ channel. At lower energies the broad distribution
corresponding to $\alpha_{1}$ and its secondary particles is
present.
Peaks from
the elastic scattering of the proton beam off boron, carbon, and small
impurities of an element close to iron can also be seen at low
energies. To select the $\alpha_{0}$ channel, a cut is placed on the
$\alpha_{0}$ peak and the CM angles projected out.
A selection of the resulting angular distribution can be seen in \cref{fig:angular-dist}.
The distributions evolve slowly from forward focussed to approximately symmetric to strongly
backward focussed.
During the analysis, it was observed that some pixels in the forward detectors
were partially shadowed by the target ladder. Conservatively these pixels were excluded from
the analysis and simulation.

In order to determine the solid angle of the detection system, a Monte
Carlo simulation was performed with the SimX
tool~\cite{Munch2018simx}. A beam of protons with a
$1\si{\mm}\times1$\si{\mm} profile was generated and propagated to a random
depth in the boron layer. At this location, an $\alpha_{0}$ was
generated and emitted isotropically in the CM system. The $\alpha_{0}$ was
propagated out of the target and into the detectors in a straight-line
trajectory. Energy losses were calculated using the SRIM tables which
approximately account for straggling~\cite{SRIM}.  The structure of
the simulation output was identical to the real data and the simulated
data was subjected to the same analysis as the real data. The
solid angle for a given CM angle can then be calculated as the ratio
between the number of detected hits to the number of simulated events. The
angular distribution is then given by the ratio of the data and
simulation for each CM angle. The differential cross section was in
this way determined by scaling the angular distribution with the integrated
current and acquisition deadtime.
The resulting angular resolved cross sections were then fitted with
the lowest five Legendre polynomials
\begin{equation}
  \label{eq:angular}
  \frac{d\sigma}{d\Omega}(\theta) = \frac{\sigma}{4\pi}\Bigg[1+\sum_{i=1}^{4}a_{i}P_{i}(\cos\theta)\Bigg].
\end{equation}

%% file: IFA009-discussion.tex
\section{Results and discussion}
\label{sec:discussion}

\begin{figure}[b]
  \centering
  \includegraphics[width=\columnwidth]{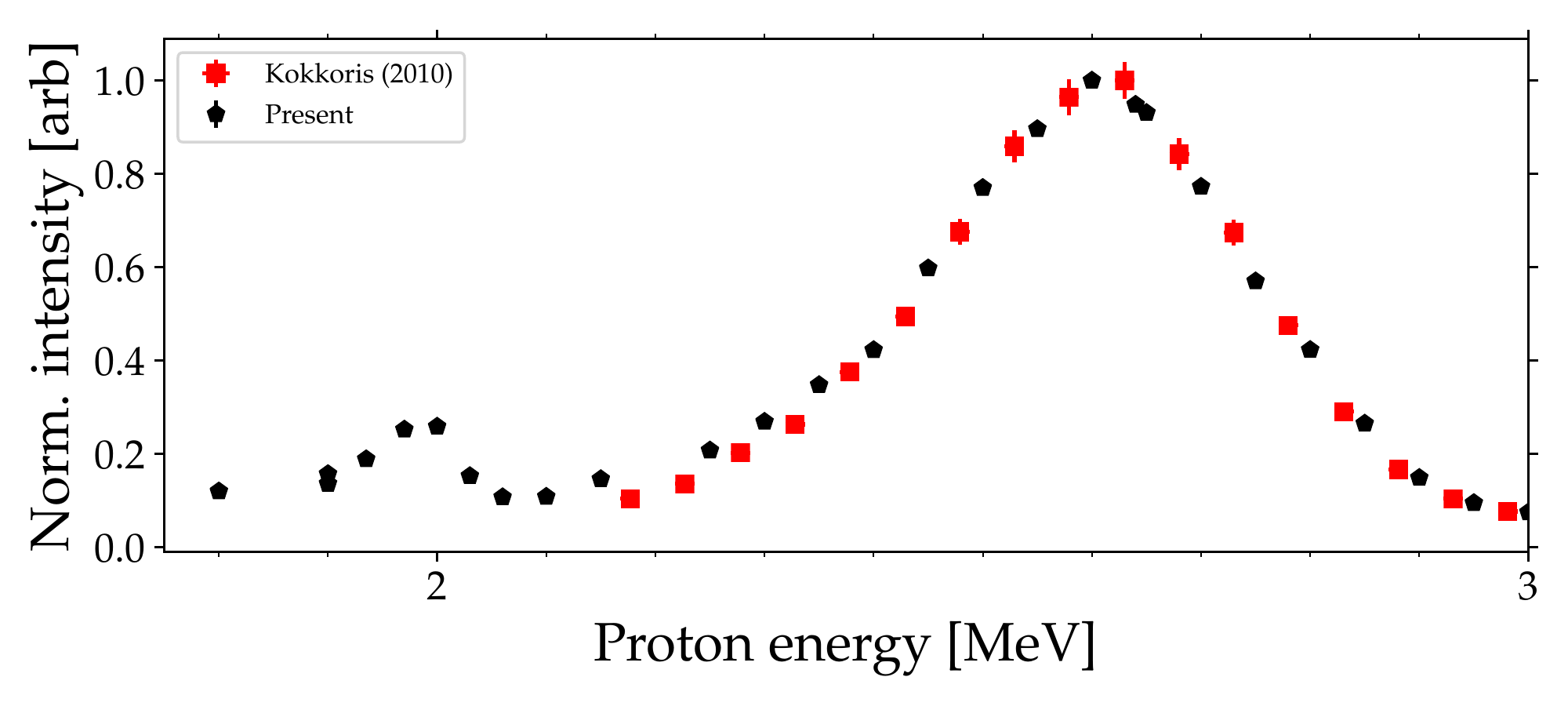}
  \caption{Comparison of the present data and the recent high-resolution dataset by
    Kokkoris \etal (KO) for $\theta=\SI{150}{\degree}$~\cite{Kokkoris2010}. Each dataset
    has been normalized so the maximum value is 1. The energy scale of the two dataset agrees
    better than \SI{5}{\keV}.}
  \label{fig:kokkoris}
\end{figure}

\begin{figure}[b!]
  \centering
  \includegraphics[width=\columnwidth,trim={0 0 10mm 0},clip]{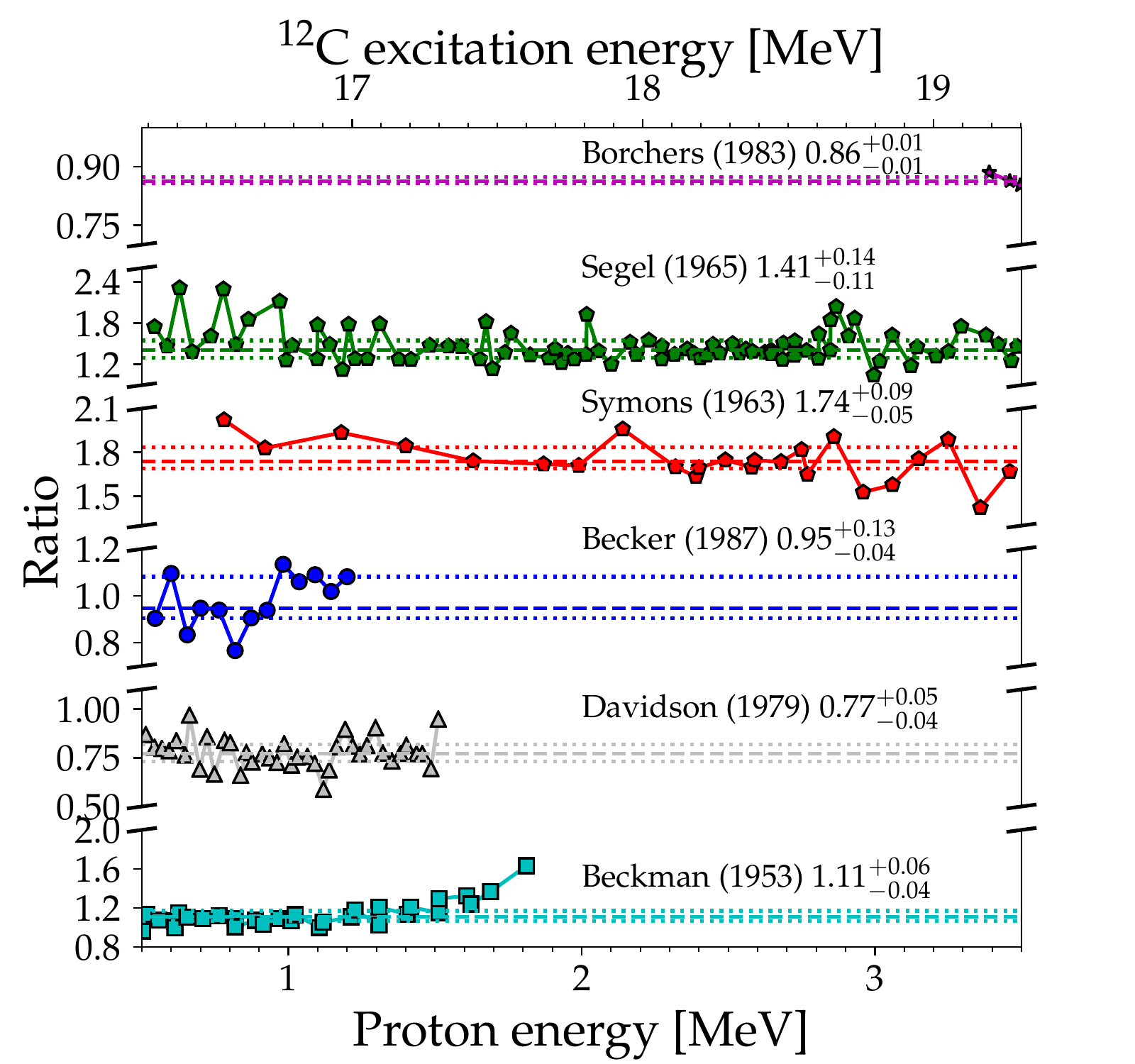}
  \caption{ Ratio of
    datasets~\cite{Beckman1953,Symons1963,Segel1965,Davidson1979,Borchers1983,Becker1987} to a
    linear interpolation of the present dataset. The dashed line
    indicates the median, while the two dotted lines indicate the 25 and 75 quartiles. See the
    text for details.  }
  \label{fig:ratio}
\end{figure}

\begin{figure}[b]
  \centering
  \includegraphics[width=\columnwidth]{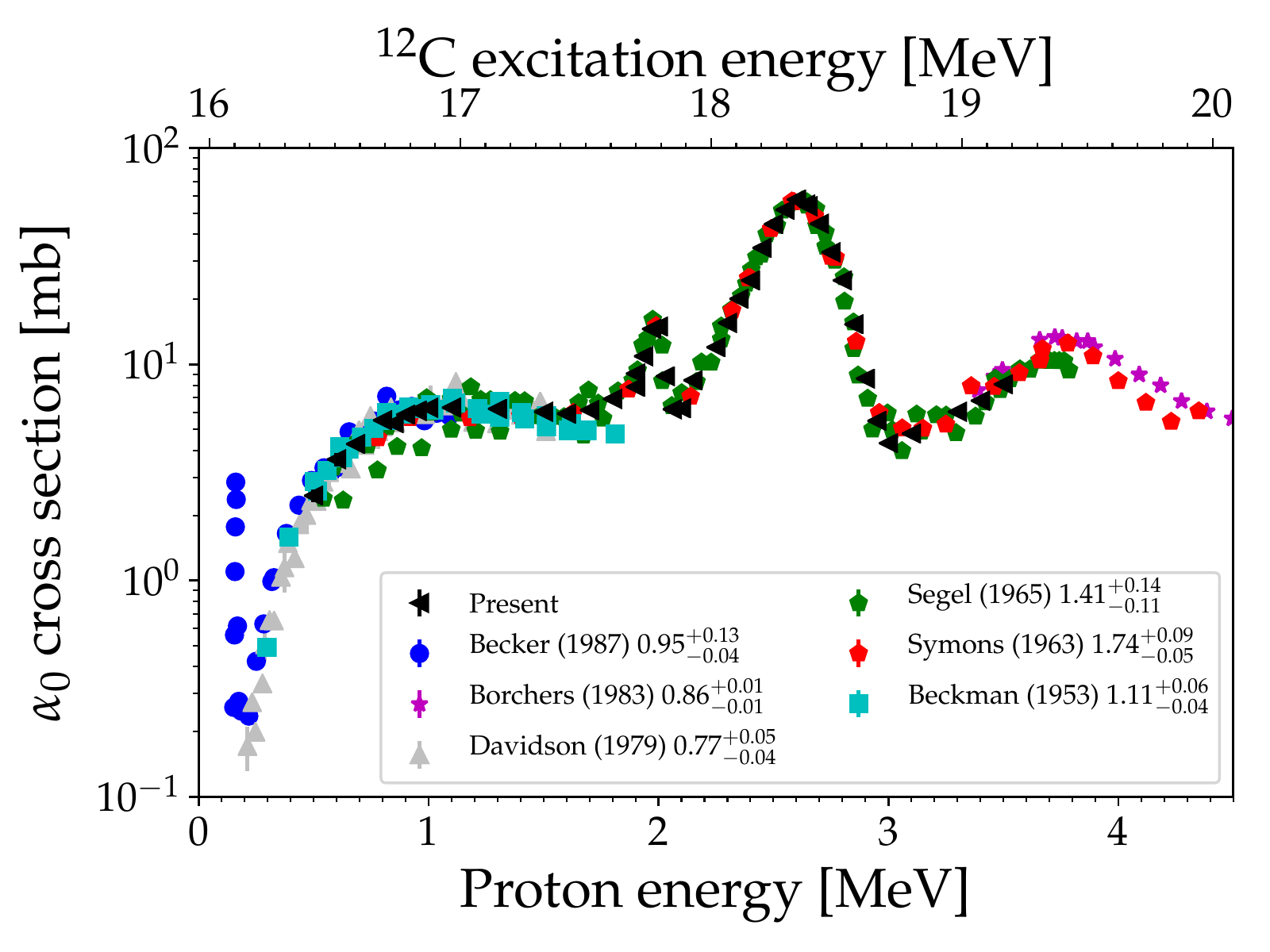}
  \caption{Adjusted $\alpha_{0}$ cross section (tabulated data available via
    EXFOR~\cite{OTUKA2014272}).
    All datasets except BE and BO have been scaled by their
    median ratio to the present dataset.}
  \label{fig:cross-adjust}
\end{figure}

The Legendre coefficients as a function of beam energy are shown in the lower four panels
of~\cref{fig:as} together with the coefficients obtained by
Refs.~\cite{Symons1963,Segel1965,Becker1987}. In addition, the position of the natural parity
states, listed in the latest compilation \cite{Kelley2017}, are also indicated.
For the two lowest-order terms, there is generally good
agreement between all four datasets in the regions of overlap. For the third-order term,
there is excellent agreement between the present data and Symons \etal (SY) while
the agreement with Segel \etal (SE) is good above $\sim \SI{1.5}{\MeV}$,
and a systematic deviation is observed below that energy.  This deviation might be due to the
experimental difficulties noted by SE for measurements below \SI{1}{\MeV}
beam energy. For the fourth-order term, there is again good agreement with SY, but a systematic
shift is seen at all energies when comparing to SE. We speculate that this is due to the latter
only measuring at six different angles whereas SY measured either 8 or 13 angles. From the data,
it appears that the $0^{+}$, $3^{-}$ and $2^{+}$ levels have an appreciable effect on the Legendre
coefficients, while the effect of the $1^{-}$ levels appear negligible. This might simply be due
to their width being as large as $\unsim\SI{1}{\MeV}$. With the present agreement with previous datasets, we are
unable to discern between the two differing interpretations presented in
Refs. \cite{Barker2002,Symons1963,Segel1965} for the region below \SI{2}{\MeV}. The reader is thus referred to them for a detailed discussion of the angular coefficients.

The angle-integrated cross section can be seen in the first panel of \cref{fig:as} and in
tabular form in \cref{tab:cross}. It shows the same
behavior as observed in the previous measurements. In \cref{fig:kokkoris} we compare data from
the present experiment and the recent high-resolution differential cross section dataset
of Kokkoris \etal (KO) for $\theta=\SI{150}{\degree}$~\cite{Kokkoris2010}.
The agreement is better than \SI{5}{\keV} which validates the overall energy calibration of the
present dataset. A similar comparison shows good agreement between the present data and that of
SY. Thus, we cannot verify the claim of KO of a discrepancy between the energy scales of KO
and SY.  However, the agreement between our data and those of SE requires the energy of SE to
be shifted down by approximately \SI{38}{\keV}.

As discussed in the introduction, there are several normalization
issues between the different
datasets~\cite{Beckman1953,Symons1963,Segel1965,Davidson1979,Borchers1983,Becker1987}. In
order to shed light on this, the ratios between the cross sections of
other datasets and the present have been computed. This can be seen in
\cref{fig:ratio} where the median ratio and the 25--75 inter quartile
ratio are written in the legend. Note that the spectrum of SE has been
shifted in energy as discussed above. The first conclusion is that
within error the present measurement is consistent with the
measurement of Becker \etal (BE), while the difference
compared to Davidson \etal is a constant scaling
factor.
There also seems to be simple scaling relations for the data of Beckman \etal below
$\unsim \SI{1.3}{\MeV}$
and the data of SE and SY in the range 1.5--\SI{2.75}{\MeV}, while larger deviations occur
outside these ranges.
The present data only overlaps with the energy range measured by  
B\"orchers \etal (BO) in a limited range.
However, within that range, the agreement is better than $2\sigma$.

Scaling all datasets except BE and BO with constant factors given in
\cref{fig:ratio} yields the cross sections seen in
\cref{fig:cross-adjust}, which confirms visually that the different
data sets can be brought into agreement by a constant
scaling. Considering the good agreement between our measurement and
the two most recent experiments by BE and BO, we argue that the overall
normalization is now better than \SI{15}{\%}.

%% file: IFA009-conclusion.tex
\section{Conclusion}
\label{sec:Conclusion}
We have measured the $\Boron(p,\alpha_0)$ reaction for proton energies
in the range \SIrange{0.5}{3.5}{\MeV}. From an analysis of the angular
distributions, total cross sections and energy dependent Legendre
polynomial coefficients have been extracted from the
data. Disagreements between previous measurements
\cite{Beckman1953,Symons1963,Segel1965,Davidson1979,Borchers1983,Becker1987} on both the
magnitude and energy dependence of the cross section have been
resolved. 
 
There is disagreement in the 
literature on the interpretation of the cross section and Legendre 
coefficients below \SI{2}{\MeV} in terms of resonances in 
\Carbon. Symons \etal interprets this as ghosts of 
resonances below the proton threshold with spin and parity $2^+$ or 
$3^-$ \cite{Symons1963}, while Barker interprets this region as a broad 
$1^-$ resonance in addition to the high energy tail of a broad $0^+$
resonance below the proton threshold in \Carbon~\cite{Barker2002}. 

We were unable to resolve this issue with a measurement of the $\alpha_{0}$ channel.
Instead this issue requires a measurement of the $\alpha_{1}$ channel and
probably also of the phase-space (Dalitz)  distributions of the 3$\alpha$
emission of that channel. That will provide additional information on
the spin and parity of the involved resonances in \Carbon, which
should be fed into a new R--matrix analysis similar to that performed
by Barker.

%% file: IFA009-acknowledge.tex
\section*{Acknowledgement}
\label{sec:Acknowledgement}

We would like to thank Folmer Lyckegaard for manufacturing the target. We
also acknowledge financial support from the European Research Council under ERC starting grant
LOBENA, No. 307447. OSK acknowledges support from the Villum Foundation through Project No.\ 10117. MM acknowledges
support from the Nustar Colaboration.